\documentclass[sigconf, screen]{acmart}
\AtBeginDocument{%
  \providecommand\BibTeX{{%
    \normalfont B\kern-0.5em{\scshape i\kern-0.25em b}\kern-0.8em\TeX}}}

\usepackage{tikz}

\setcopyright{none}
\renewcommand\footnotetextcopyrightpermission[1]{}
\acmConference[]{}{}{}
\begin{document}

\title{Story Designer: Towards a Mixed-Initiative Tool to Create Narrative Structures}

\author{Alberto Alvarez}
\affiliation{%
  \institution{Malmö University, Game Lab}
  \city{Malmö}
  \country{Sweden}
  \postcode{21119}
}
  \email{alberto.alvarez@mau.se}

\author{Jose Font}
\affiliation{%
  \institution{Malmö University, Game Lab}
  \city{Malmö}
  \country{Sweden}
  \postcode{21119}
}
\email{jose.font@mau.se}

\author{Julian Togelius}
\affiliation{%
  \institution{New York University}
  \city{New York}
  \country{USA}}
\email{julian@togelius.com}

\renewcommand{\shortauthors}{Alvarez. et al}

\begin{abstract}
Narratives are a predominant part of games, and their design poses challenges when identifying, encoding, interpreting, evaluating, and generating them. One way to address this would be to approach narrative design in a more abstract layer, such as narrative structures. This paper presents Story Designer, a mixed-initiative co-creative narrative structure tool built on top of the Evolutionary Dungeon Designer (EDD) that uses tropes, narrative conventions found across many media types, to design these structures. Story Designer uses tropes as building blocks for narrative designers to compose complete narrative structures by interconnecting them in graph structures called narrative graphs. Our mixed-initiative approach lets designers manually create their narrative graphs and feeds an underlying evolutionary algorithm with those, creating quality-diverse suggestions using MAP-Elites. Suggestions are visually represented for designers to compare and evaluate and can then be incorporated into the design for further manual editions. At the same time, we use the levels designed within EDD as constraints for the narrative structure, intertwining both level design and narrative. We evaluate the impact of these constraints and the system's adaptability and expressiveness, resulting in a potential tool to create narrative structures combining level design aspects with narrative.
\end{abstract}



\begin{teaserfigure}
\centering
  \includegraphics[width=16cm]{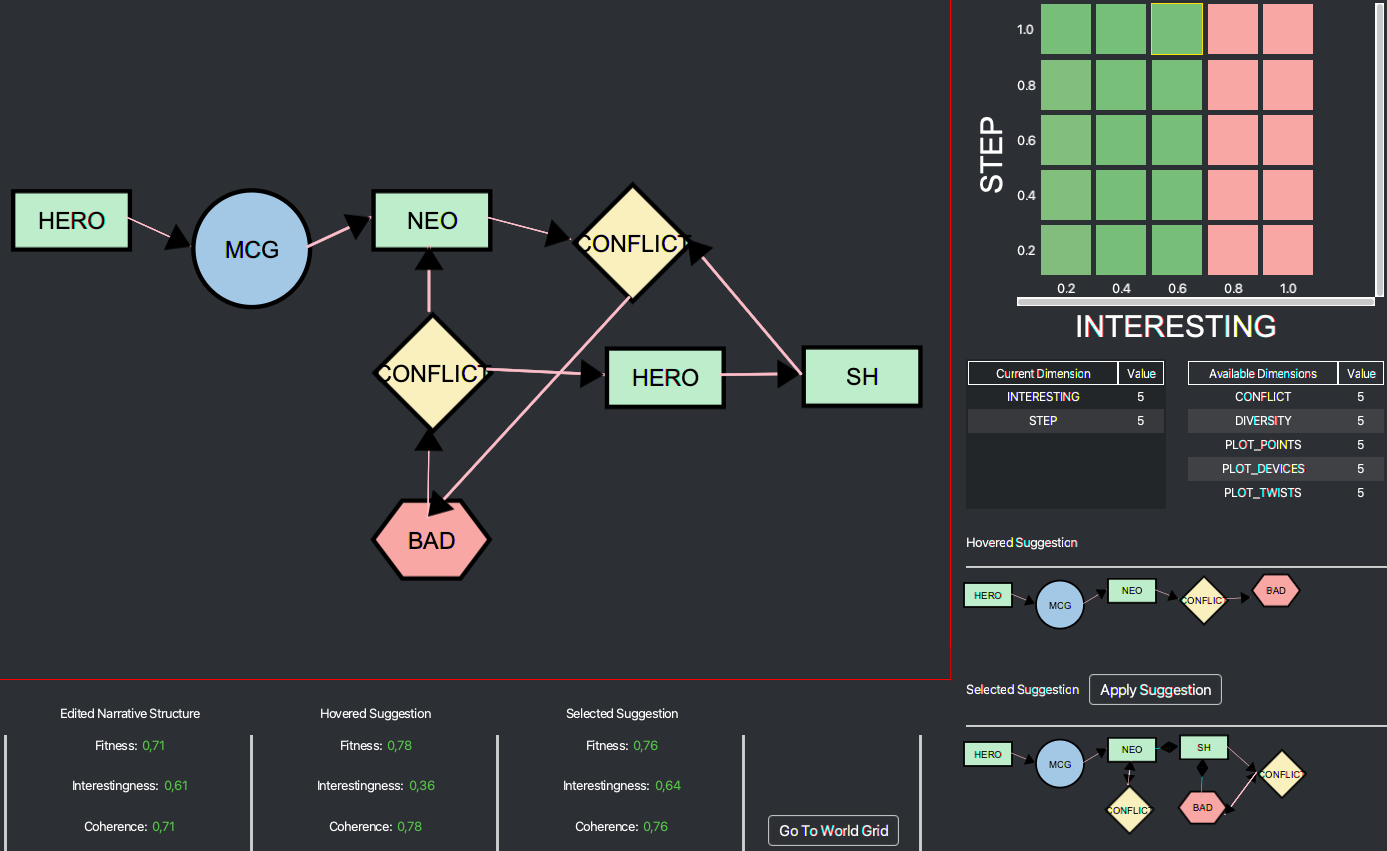}
\caption{The Story Designer screen in the Evolutionary Dungeon Designer. In the center, there is the main narrative graph being edited by the designer. to the right, the suggestion grid using the Interactive Constrained MAP-Elites (IC MAP-Elites), the possible dimensions to be used, and two inspected suggestions. At the bottom, Story Designer presents some extra information regarding fitness, interestingness, and coherence, for the designer's convenience.}
    \label{fig:story-screen}
\end{teaserfigure}

\maketitle

\section{Introduction}




Games are multifaceted content that intertwines gameplay, mechanics, audio, level, graphics, and narrative facets~\cite{liapis_orchestrating_2019}. Narrative has been linked as a key facet to connect different components in games such as level design~\cite{kishino_hunt_2005}, and to create meaningful interactions with depth and context~\cite{ashmore_quest_2007,kybartas_quinn_survey_2017}. Thus, narrative in games has been a focus of study~\cite{aarseth_narrative_2012,eladhari_interweaving_2014,yu_what_2020} and its generation has been approached in different ways and with different techniques~\cite{smith_situating_2011,ammanabrolu_story_2020,green_data_2018,tambwekar_controllable_2019,alvarez_questgram_2021}.

Patterns are a common approach to narrative and other facets. The focus then has been on extracting common narrative aspects to ease the identification, encoding, and generation of narrative in different forms~\cite{doran_prototype_2011,alvarez_tropetwist_2022,trenton_quest_2010,treanor_ai-based_2015,rouse_sketching_2018,breault_let_2021}. However, it remains a challenge to define certain narrative aspects more aligned with the structure and overarching goals of the game given what type of content is generated, as well as using these to design and compare among games. One approach would be to change the abstraction level at which the narrative is designed. Instead of focusing on the details, quests, or plot, one could focus on the structure. Narrative structures can be used to describe how a story is to be developed, as argued by Barthes~\cite{barthes_introduction_1966}, and to create an abstract representation that reveals common structures among them, such as Propp's 31 ``narremes" ~\cite{propp_morphology_1975}. One approach to generate narrative structures is TropeTwist~\cite{alvarez_tropetwist_2022}, which uses tropes, narrative conventions found across many media types~\cite{lewis_governing_2018,garcia-sanchez_simpsons_2021}, as patterns to design these structures.

This paper presents Story Designer, a mixed-initiative co-creative (MI-CC) narrative structure tool built on top of the Evolutionary Dungeon Designer (EDD) using the TropeTwist system. Story Designer uses tropes as building blocks for designers to compose complete narrative structures by interconnecting them in graph structures called narrative graphs. Story Designer lets designers create narrative graphs and assist them with a suggestion grid that uses the Interactive Constrained MAP-Elites (IC MAP-Elites)~\cite{alvarez_interactive_2020}. By having an MI-CC system to design narrative structures, designers could ideate and prototype their structures while the system adapts and suggest novel narratives, making use of patterns, optimizing coherence, and situating the narrative structures along dimensions of interest for designers. At the same time, IC MAP-Elites can take advantage and use the designer's structure as a proxy to evaluate subjective characteristics such as interestingness, which has been the subject of several studies~\cite{perez_model_2013,lankoski_models_2013,rowe_storyeval_2009}.


As Story Designer is implemented in EDD and based on the link between level design and narrative, we make use of the designed dungeon to create constraints over the narrative generation, effectively intertwining both facets. We assess Story Designer with four controlled and simulated experiments, three premade structures of different games, and one step by step design that showcase the possibilities within the system. All experiments were tested with and without level design constraints and using a pair of dimensions and all dimensions during the search. Our results indicate that IC MAP-Elites have consistency and stability in generating content and that delimiting the search space with additional level constraints, while limiting the diversity and generation of complex structures, guides better the search.
\section{Related Work}



There is by now a large body of research on procedurally generating various types of game content~\cite{shaker_procedural_2016}. While the literature on PCG in general is far too voluminous to survey here, it should be noted that PCG methods of different kinds have been developed for a wide variety of content, not just game levels. Narrative, quests, and plots have been generated using different approaches such as planning~\cite{young_plans_2013}, grammars~\cite{hartsook_toward_2011,doran_prototype_2011}, machine learning~\cite{tambwekar_controllable_2019}, and patterns~\cite{trenton_quest_2010,breault_let_2021,alvarez_tropetwist_2022}. Further, several approaches have been proposed to generate multiple facets of games, in particular level geometry together with rules, music, lighting, sound etc~\cite{liapis_computational_2014,liapis_orchestrating_2019,hoover_audioinspace_2015,holtar_audioverdrive_2013,karavolos_multi-faceted_2019,cook_rogue_2014,treanor_game-o-matic_2012,green_data_2018}. More relevantly to the current project, several papers have proposed ways of co-generating narrative and levels~\cite{ashmore_quest_2007,hartsook_toward_2011,dormans_generating_2011,abuzuraiq_taksim_2019}.


In tandem with research on automatically and autonomously generating game content and narrative, there has been a considerable amount of work ``mixed-initiative'' systems, which allow a human designer to co-create content with algorithms. In the domain of level generation for games, a number of systems have been developed that allow a human to receive suggestions, feedback, or constraints from an AI systems. These include systems for co-creating platform game levels~\cite{smith_tanagra_2011}, strategy maps~\cite{liapis_sentient_2013}, and certain aspects of narrative~\cite{kreminski_germinate_2020,kreminski_why_2020}.

The core algorithm employed in the current paper is MAP-Elites, a quality-diversity algorithm that seeks to illuminate a space of possible problem solutions~\cite{mouret_illuminating_2015}. While essentially a type of evolutionary algorithms, MAP-Elites, like other quality-diversity algorithms, do not seek to find a single best solution but rather a set of solutions that vary along certain specified measures. The measures define a grid, where each cell is the best solution that has been found within certain values of the measures. These measures can be defined in many ways; for game levels, they might include the density of a level, its difficulty for a particular type of agent, its symmetry etc. MAP-Elites has been used in multiple recent AI-based game design systems~\cite{alvarez_assessing_2021,alvarez_empowering_2019,charity_baba_2020,charity_mech-elites_2020,khalifa_talakat_2018}.
\section{Story Designer}

Story Designer is a new system integrated in EDD, which presents a visual interface for mixed-initiative narrative structure generation. It makes extensive use of the TropeTwist system as foundation to build narrative graphs and assess them by identifying trope patterns. The user manually designs a story structure by adding and interconnecting nodes in a graph, which seeds an evolutionary algorithm (EA) that generates story structure suggestions that can be incorporated into the user's design. This continuous co-creative design process implements the Interactive Constrained MAP-Elites (IC MAP-Elites) approach presented in~\cite{alvarez_empowering_2019}, providing quality-diverse suggestions across several feature-dimensions.

Story Designer is interconnected with the level design facet in EDD. This means that the narrative graphs that can be developed and that can be generated and suggested are constrained by the content that exists in the levels. For instance, if the designer adds two NPCs besides the Hero, then the system could at most, use three character nodes to represent them, or if the designer adds a boss enemy and a quest item, this would mean that the boss enemy could be represented as one of the villain nodes (e.g., Enemy, Big Bad, or Dragon) and the quest item as a possible Plot Device.

\subsection{TropeTwist}

TropeTwist~\cite{alvarez_tropetwist_2022} is a system that uses tropes~\cite{lewis_governing_2018,garcia-sanchez_simpsons_2021,richmond_tv_2004,harris_periodic_2016}, narrative conventions easily recognizable by the audience, as patterns that combine to compose narrative structures. These structures define generic aspects of a story, leading to the identification of events, roles, and other relevant narrative elements arranged as nodes in an interconnected narrative graph. By having all this elements in a graph, entire narratives are encoded using graph grammars, to then procedurally generate novel narrative variations by means of a MAP-Elites algorithm that considers several narrative evaluation metrics, such as interestingness, coherence, and cohesion. 

Nodes in a narrative graph represent tropes. Interconnected tropes create other composite tropes and patterns, that can be identified as subgraphs of a complete narrative graph. These patterns can be \textbf{micro-patterns} encapsulating a single trope node, \textbf{meso-patterns}, often composed by more than one micro-pattern with a specific meaning, and \textbf{auxiliary patterns}, identifying structural gaps in the graph. For a detailed definition of all tropes and patterns, please refer to~\cite{alvarez_tropetwist_2022}. Here we present a comprehensive summary:

\begin{itemize}
    \item Micro-patterns are the fundamental narrative unit in the system, encapsulating tropes in building blocks to create complex narrative structures. These are classified into structure patterns (SP), that articulate the story elements (i.e. Conflict), character patterns (CP) (i.e. heroes and villains), and plot device patterns (PDP), that move the story forwards towards a particular goal (i.e. the MacGuffin).
    \item Meso-patterns may emerge from the combination of micro-patterns and other meso-patterns, denoting spatial, semantic, and usability relationship within the narrative graph.
    \begin{enumerate}
        \item The \emph{Conflict Pattern (ConfP)} ties a conflict node to two other nodes representing both parties in a conflict (i.e. HERO $\rightarrow$ CONFLICT $\rightarrow$ EMP, a hero against the Empire).
        \item The \emph{Derivative Pattern (DerP)} defines relations of entailment between other nodes, called derivatives. These derivatives acquire a local and temporal order, and a causal relationship. I.e the former conflict connected to EMP $\diamondsuit$--- DRA $\diamondsuit$--- NEO, means that the hero engages the Empire, which entails both a conflict with the Dragon (\emph{DRA}) and the appearance of the Chosen One (\emph{NEO}).
        \item The \emph{Reveal Pattern (RevP)} connects two independent CPs as one, meaning that character A was, in fact, always character B, and vice-versa. This pattern turns all existing conflicts between them into \emph{fake} conflicts.
        \item The \emph{Active Plot Device Pattern (APD)} triggers a PDP and integrates it in the the narrative, since PDP are passively described and lack any start condition.
        \item \emph{Plot Points (PP)} are key discrete narrative events. The derivatives within a \textit{DerP}, the source of a reveal pattern, as well as active plot devices are considered plot points.
        \item A \emph{Plot Twist (PT)} identifies those plot points that could change the natural flow of the narrative. I.e. in EMP $\diamondsuit$--- DRA $\diamondsuit$--- NEO, NEO is identified as a plot twist since its nature (heroic) is opposed to that of the first node EMP (villainous), which alters the natural order of the connecting derivative pattern.
    \end{enumerate}
    \item Auxiliary patterns spot and encapsulate those areas in the graph that don't contain meaningful narrative information. \textit{Nothing} highlights nodes that are not identified or part of any meso-pattern; whereas \textit{Broken Link} marks outgoing connections from any node that do not lead to any pattern.
\end{itemize}

\subsection{Workflow}

Story Designer is integrated in EDD as a separate view (Figure \ref{fig:story-screen}) that can be accessed anytime from the dungeon editor. The use starts with a minimal sample narrative graph HERO $\rightarrow$ CONFLICT $\rightarrow$ ENEMY in the manual edition pane (center). This graph can be extended by adding nodes from the node context menu that pops up with a right-click on an empty space. Node are arranged by type for the sake of clarity, and an option to automatically re-arrange the graph is shown at the end of the menu. Right-clicking on an existing node border will pop up the edge context menu, that allows the user to create a new connection or to delete the selected node. Existing connections are deleted by left-clicking on them.

In a way similar to EDD's room editor \cite{alvarez_empowering_2019}, as the user edits the narrative graph manually, this graph is fed into the underlying evolutionary algorithm that procedurally generates on the fly alternative narrative graphs in the suggestions pane (right). The top-right corner shows the feature-dimension matrix, whose cells are colored depending on the fitness of the fittest elite contained in it, ranging from dark red (no elite yet), to dark green (optimal fitness). The elite in the selected cell of the matrix is displayed in the bottom-right corner. Hovering the mouse above a cell displays its elite's graph above the selected one, which allows the user to compare several graphs at a glance.

\subsection{Evolving narrative structures with Graph Grammars}

The underlying evolutionary algorithm in Story Designer is an adapted version of IC MAP-Elites~\cite{alvarez_interactive_2020} to evolve grammars. In Story Designer, an individual's phenotype is a narrative graph, and its encoding genotype is a graph grammar. A graph grammar is a context-free grammar whose productions add, remove, and modify nodes and edges to a graph. 

An individual's genotype is the production rules of the grammar, which are deterministic i.e., a production rule (or pattern) only matches one production. Given that the graph grammar does not need to be applied sequentially until terminal nodes are reached, every individual does a random sampling of the rules in their genotype to produce \emph{recipes}. \emph{Recipes} simply describe the order of rules to be applied (sequentially) and the amount of times they will be applied. \emph{Recipes} do not have repetitions within them i.e., if rule 1 is added at step 2, subsequent addition would simply add to the amount of times that rule will be applied at step 2. The internal parts of the EA works exactly as in TropeTwist, but now it is extended to use all the capabilities of IC MAP-Elites, namely, the continuous adaptive evolution aspect~\cite{alvarez_tropetwist_2022,alvarez_interactive_2020}.

Moreover, thanks to continuous evolution, the EA constantly incorporates the most recent version of the user's design to the population of individuals in the corresponding cell of the feature-dimension matrix. The designer can switch between dimensions and their granularity at any given time. IC MAP-Elites manages two different populations within each cell: a feasible and an infeasible one. Individuals move across cells when their dimension values change or between the feasible and infeasible population according to their fulfillment of the feasibility constraint. Narrative graphs are deemed infeasible if they are not fully connected (i.e., all nodes can be reached from an arbitrary starting point) and if there exists a conflict pattern within the graph with more than one self-conflict. If level design constraints are enabled, narrative graphs that violate any level design constraint are also infeasible. Infeasible individuals are evaluated (equation~\ref{eq:inf_fitness}) in a weighted sum ($w_{0}=0.5, w_{1}=0.25, w_{2}=0.25$) based on how close they are to be fully connected and to removing inadequate self-conflicts while trying to maximize the graph's cohesion.


\begin{multline}
\label{eq:inf_fitness}
f(infeasible) = w_{0} \times f(cohesion) + w_{1} \times \frac{\#!reachable_{V(NG)}}{|V(NG)|} 
\\ + w_{2} \times  \frac{\#!valid_{NG(self_conf)}}{|V(NG)|}
\end{multline}

Generated narrative graphs that are deemed feasible, are evaluated on their coherence (equation~\ref{eq:coherence_fitness}), which is used to assess how correct, coherent, and in general, syntactically correct the narrative graphs are. Coherence aims at maximizing an equally weighted sum between cohesion and consistency (eq.~\ref{eq:consistency_fitness}). Cohesion refers to the link between elements that hold together to form some group, which in Story Designer means the minimization of auxiliary patterns (\textit{Nothing} and \textit{Broken Link}) within the narrative graph. Consistency means that the narrative graph should be regular and free of contradictions, aiming at maximizing the quality of micro-patterns and minimize contradictions created by meso-patterns (contradictions can affect the consistency fitness up to $w_{0}=0.3$). For a more detailed explanation of how the EA works internally and the different fitness functions, we refer to the TropeTwist paper~\cite{alvarez_tropetwist_2022}.








\begin{equation}
\label{eq:consistency_fitness}
f_{consistency} = \frac{\sum_{i=0}^{len(ng_{micro})} i_{qual}}{len(ng_{micropat})} -  \\ 
w_{0} \times \frac{len(ng_{fakeConfP})}{len(ng_{confP})} 
\end{equation}

\begin{equation}
\label{eq:coherence_fitness}
f(coherence) = f(consistency) + (1.0 - f(cohesion))
\end{equation}

\subsection{Behavior Dimensions for Graph Grammars}

Dimensions in MAP-Elites are a key component for the search space to be delimited, and are identified as those aspects of the individuals that can be calculated in the behavioral space, and that are independent of the fitness calculation. In Story Designer, the designer is able to pick two dimensions at a time to facilitate visualization, and all dimensions, when needed, are limited using a threshold $\delta = 5$. TropeTwist implemented \textit{Interestingness} and \textit{Step} as behavior dimensions when using MAP-Elites to generate novel narrative graphs. \textit{Step} is calculated as the Levenshtein distance between two narrative graphs, taking into account the amount of nodes and connections and their type (eq. \ref{eq:StepDim}). \textit{Interestingness} make use of the APDs, Plot Points, and Plot Twists that are present in a narrative graph to assess an approximate semantic evaluation since those represent some type of variation in the graph (eq.~\ref{eq:interesting_fitness}). Given \textit{Interestingness} is a highly subjective measurement, we rely on those patterns since they calculate their quality based on the current narrative graph and the one being edited by the designer.

\begin{equation}
\label{eq:StepDim}
D_{step} =  \frac{lev_{a,b} (|a|, |b|)}{\theta}
\end{equation}

\begin{equation}
\label{eq:interesting_fitness}
D_{int} = w_{0} \times \frac{APD_{q}}{\#APD} + w_{1} \times \frac{\#PP_{q}}{\#PP} +  w_{2} \times \frac{PT_{q}}{\#PT}ng
\end{equation}

Furthermore, we have extended TropeTwist with five more dimensions relevant to the narrative structure design process, to give more choice to designers and experiment with other dimensions in the search space:



\begin{figure*}[t]
    \centering
    \includegraphics[width=\textwidth]{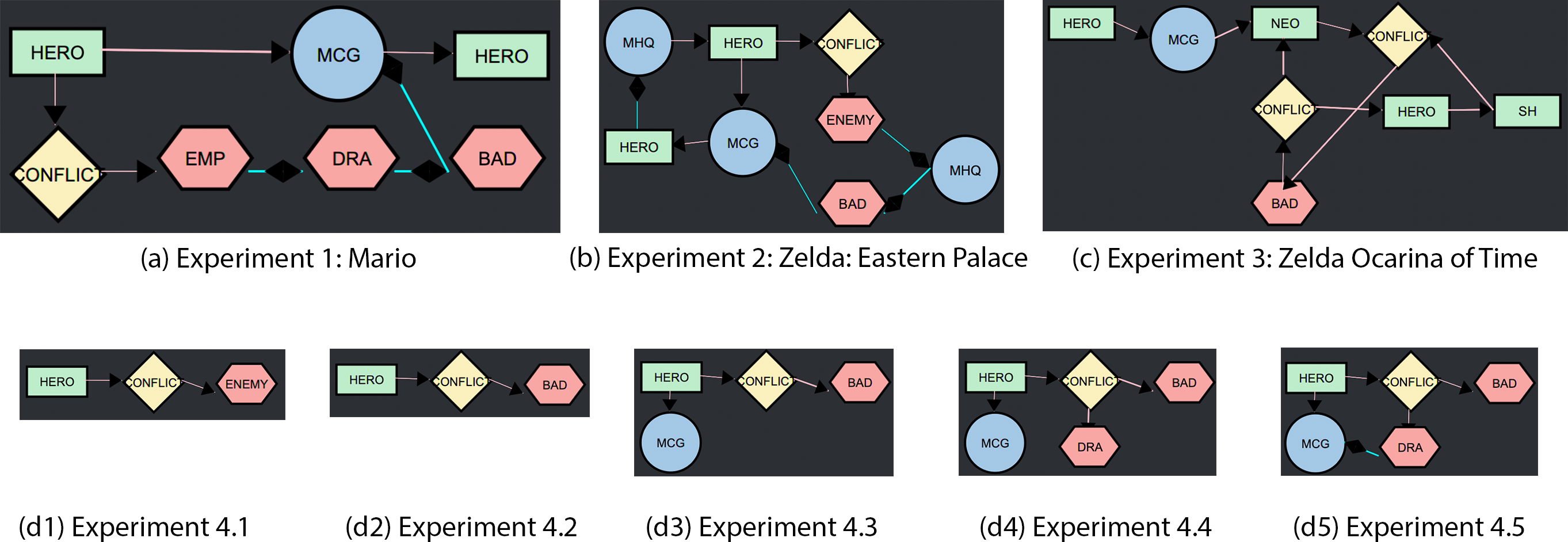}
    \caption{Narrative graphs used for the experiments, constructed and designed in Story Designer. When experiment 4 is discussed, the narrative graph referred is Experiment 4.5 as that is the design's final step.}
    \label{fig:examples}
\end{figure*}

\textbf{Diversity (div).} Diversity measures the variety of [base] trope types within a narrative structure. Currently, there exist four base trope types, \textit{Hero} (h), \textit{Villain} (v), \textit{Structure} (s), \textit{Plot Devices} (pd). Diversity takes into account the tropes that also extend these base tropes. Thus, $D_{div}$ collects all tropes within a graph, and increase a counter for each of the base trope type ($NG_{base} = h, v, s, pd \in NG$), normalized by the max amount of base trope types depicted in Eq.~\ref{eq:diversityDim}:


\begin{equation}
\label{eq:diversityDim}
D_{div} =  \frac{NG_{base}}{\#Trope_{base}}
\end{equation}


\textbf{Conflict (confs).} Since we already calculate all patterns within a narrative graph, conflict simply calculates the amount of \emph{explicit} conflict patterns ($\#NG_{c} = C_{exp} \in allPatterns$) that exist within a narrative graph normalized by a conflict threshold $\omega = 5$. We use $\omega$ to avoid stimulating the generation of narrative graphs with a massive amount of conflicts, which could create noise in the evolution and focus on the conflicts rather than other tropes and patterns. $D_{confs}$ is then calculated as $\frac{\#NG_{c}}{\omega}$.



\textbf{Plot points (pp).} Plot points measures the amount of plot points within a narrative graph ($\#NG_{pp} = pp \in allPatterns$) and normalize it by $\delta$. Given that plot points are dynamically assessed based on other patterns and combination of tropes, we limit the dimension with $\delta$ to avoid losing coherence in favor of generating more plot points. $D_{pp}$ is calculated as $\frac{\#NG_{pp}}{\delta}$.

\textbf{Plot Twist (pt).} Plot twist measures the amount of plot twists within a narrative graph ($\#NG_{pt} = pt \in allPatterns$) and normalize it by $\delta$. Plot twists relate to special situations within a narrative graph where the somewhat abrupt change in the tropes or combination of tropes could alter the narrative and create a surprise effect. Therefore, we limit the dimension with $\delta$ to avoid ``degenerating" narratives with too many twists. $D_{pp}$ is calculated as $\frac{\#NG_{pt}}{\delta}$.

\textbf{Plot devices (pd).} Plot devices measure the amount of active plot devices within a narrative graph ($\#NG_{pd} = apd \in allPatterns$) and normalize it by $\delta$. Plot devices create targets and goals within a narrative, and active plot devices operationalize these in the narrative graph associating them with multiple tropes; thus, similar to \emph{pt}, we limit with $\delta$ to avoid ``degenerating" the narrative. $D_{pd}$ is calculated as $\frac{\#NG_{pd}}{\delta}$.

\section{Experiment Setup}

\begin{table}[]
\caption{Level constraints used per Experiment. Constraints were chosen based on the maximum amount of elements needed to design the narrative structure in the system.}
\begin{tabular}{l|llll}
Constraining elements & Exp. 1 & Exp. 2 & Exp. 3 & Exp. 4 \\ \hline
Heroes                & 2            & 2            & 4            & 2            \\
Enemies               & 2            & 2            & 1            & 2            \\
Quest Items           & 2            & 3            & 1            & 2           
\end{tabular}
\label{tab:level-constraints}
\end{table}
\begin{table*}[]
\caption{Average values from the experiments using Interestingness and Step as dimensions. Values in bold represent the best values in the specific experiment between using or not using level constraints. $\star$ represents the best values across experiments within their specific condition (using or not level constraints).}
\label{tab:exp-int-step}
\resizebox{\textwidth}{!}{%
\begin{tabular}{|l|llll|llll|}
\hline
               & \multicolumn{4}{c|}{With level constraint}                                             & \multicolumn{4}{c|}{No level constraint}                                                                 \\ \hline
Experiment     & Avg. coverage     & Avg. Uniques      & Avg. fitness              & Avg. Int           & Avg. coverage              & Avg. Uniques               & Avg. fitness       & Avg. Int                  \\ \hline
Experiment 1   & 20.7\%±2.8        & 132±37.6          & \textbf{0.82±0.01$\star$} & 0.35±0.02          & \textbf{24.2\%±2.2}        & \textbf{209.6±27.8}        & 0.8±0.01$\star$    & \textbf{0.36±0.03}        \\
Experiment 2   & 19.1\%±0.9        & 161.8±20.2        & \textbf{0.82±0.02}        & 0.37±0.02          & \textbf{21.4\%±2}          & \textbf{290.6±27.2$\star$} & 0.8±0.01$\star$    & \textbf{0.37±0.01}        \\
Experiment 3   & 20.6\%±3.9        & 173.6±55.5$\star$ & \textbf{0.78±0.02}        & 0.39±0.03$\star$   & \textbf{24.9\%±1.9}        & \textbf{244.2±15.9}        & 0.76±0.02          & \textbf{0.39±0.02$\star$} \\
Experiment 4   & 22.5\%±4.1$\star$ & 156.4±53.6        & 0.8±0.04                  & \textbf{0.34±0.02} & \textbf{27.8\%±2.2$\star$} & \textbf{267.8±30.3}        & \textbf{0.8±0.02}  & 0.33±0.03                 \\ \hline
Experiment 4.1 & 6.5\%±3           & 49±31.9           & 0.67±0.09                 & 0.07±0.02          & \textbf{9.8\%±2.6}         & \textbf{83.8±40.3}         & \textbf{0.72±0.04} & \textbf{0.13±0.05}        \\
Experiment 4.2 & 4.3\%±2.7         & 23.8±16.7         & 0.68±0.07                 & 0.12±0.08          & \textbf{6.8\%±2.6}         & \textbf{52±28.2}           & \textbf{0.75±0.04} & \textbf{0.15±0.05}        \\
Experiment 4.3 & 10.9\%±3.7        & 62.4±26.9         & 0.75±0.03                 & 0.27±0.04          & \textbf{15.4\%±5.4}        & \textbf{124±52.9}          & \textbf{0.79±0.03} & \textbf{0.32±0.05}        \\
Experiment 4.4 & \textbf{15\%±2.1} & 94.8±19.5         & 0.79±0.04                 & 0.29±0.06          & 14.9\%±3.8                 & \textbf{116.2±41}          & \textbf{0.85±0.03} & \textbf{0.35±0.05}        \\
Experiment 4.5 & 14.1\%±4.6        & 82.8±37.9         & 0.83±0.04                 & \textbf{0.33±0.03} & \textbf{14.6\%±3.2}        & \textbf{93.2±36.9}         & \textbf{0.85±0.02} & 0.32±0.04                 \\ \hline
\end{tabular}%
}
\end{table*}
\begin{table*}[]
\caption{Experiments using all possible dimensions (7 dimensions) as behavioral dimensions in the MAP-Elites search. Coverage relates to the pair Interestingness-Step for comparison with Study 1. Values in bold represent the best values in the specific experiment between using or not using level constraints. $\star$ represents the best values across experiments within their specific condition (using or not level constraints).}
\label{tab:exp-all-dims}
\resizebox{\textwidth}{!}{%
\begin{tabular}{|l|llll|llll|}
\hline
               & \multicolumn{4}{c|}{With level constraint}                                       & \multicolumn{4}{c|}{No level constraint}                                                                       \\ \hline
Experiment     & Avg. coverage     & Avg. Uniques       & Avg. fitness       & Avg. Int           & Avg. coverage              & Avg. Uniques              & Avg. fitness              & Avg. Int                  \\ \hline
Experiment 1   & 33.2\%±1.5        & 496.2±132.3        & 0.75±0.05$\star$   & 0.32±0.03          & \textbf{36.7\%±1.8}        & \textbf{1257.8±165.3}     & \textbf{0.75±0.02$\star$} & \textbf{0.33±0.02}        \\
Experiment 2   & \textbf{29\%±3.4} & 760.8±100.9$\star$ & \textbf{0.72±0.01} & 0.32±0.02          & 27.8\%±2                   & \textbf{1221.8±229.3}     & \textbf{0.72±0.01}        & \textbf{0.33±0.01}        \\
Experiment 3   & 33.6\%±1.5        & 658.2±48.2         & \textbf{0.71±0.01} & 0.34±0.01$\star$   & \textbf{37.2\%±1.7}        & \textbf{1357.4±48$\star$} & 0.68±0.01                 & \textbf{0.36±0.01$\star$} \\
Experiment 4   & 35.2\%±2$\star$   & 690.4±325.4        & 0.71±0.03          & \textbf{0.29±0.02} & \textbf{38.3\%±1.8$\star$} & \textbf{1314.6±181.1}     & \textbf{0.73±0.01}        & 0.28±0.02                 \\ \hline
Experiment 4.1 & 6.5\%±1.4         & 56.8±8.5           & \textbf{0.67±0.06} & 0.04±0.02          & \textbf{10.2\%±1.5}        & \textbf{118.6±43.5}       & 0.65±0.06                 & \textbf{0.07±0.03}        \\
Experiment 4.2 & 8.8\%±1.9         & 74.8±19.9          & \textbf{0.59±0.04} & 0.07±0.02          & \textbf{9.9\%±2.5}         & \textbf{153.6±21.9}       & 0.56±0.1                  & \textbf{0.09±0.05}        \\
Experiment 4.3 & 16.4\%±2.1        & 123.8±17.7         & \textbf{0.62±0.04} & 0.19±0.02          & \textbf{23\%±1.5}          & \textbf{377.2±81.2}       & 0.61±0.07                 & \textbf{0.25±0.02}        \\
Experiment 4.4 & 17.8\%±3.5        & 155.8±25.5         & 0.65±0.06          & 0.21±0.05          & \textbf{25.8\%±1.5}        & \textbf{286.8±209.3}      & \textbf{0.65±0.05}        & \textbf{0.26±0.01}        \\
Experiment 4.5 & 23\%±3.2          & 232.8±69.8         & \textbf{0.67±0.02} & 0.28±0.03          & \textbf{30.8\%±3}          & \textbf{891.8±202.3}      & 0.67±0.04                 & \textbf{0.3±0.01}         \\ \hline
\end{tabular}%
}
\end{table*}


We ran a set of experiments using different narrative graphs as starting points and level constraints to evaluate Story Designer, the use of TropeTwist with IC MAP-Elites, and its adaptability. Our goal is to analyze how IC MAP-Elites can adapt to the designer's narrative graph and how that affects the search space. At the same time, we explore how we can connect level and narrative and the effect using level design constraints have on the development and generation of narrative structures.

We ran each experiment $5$ times, set the initial population to $1000$ randomly created grammars, and each individual is limited to test $5$ recipes regardless of the chromosome size. Offspring were produced either by selecting either the left-side or right-side of a random production rule and exchanging them or with a $50$\% mutation chance. If an offspring was generated by mutation, there was a $10$\% chance to add or remove a production rule and a $90$\% to modify existing production rules in various ways (e.g., removing, adding, or changing part the rule). When using level constraints, these were enforced as feasibility constraints, effectively setting individuals as infeasible if violating any constraint.

For each experiment, we used the dimension pairs \textbf{Interestingness}\textbf{-Step} and all dimensions for IC MAP-Elites to compare different space constraints that would be employed in an MI-CC system and a full space search across dimensions similar to Alvarez et al.'s work~\cite{alvarez_interactive_2020}. Experiments 1-3 consist of reusing the three proof-of-concept narrative graphs used by Alvarez and Font~\cite{alvarez_tropetwist_2022}, and testing them in Story Designer; assessing the impact of level constraints and how the space is explored in comparison with previous results. Experiment 4 assesses the same as experiments 1-3 but focuses on evaluating the system's adaptability and how IC MAP-Elites respond to design changes, which enables different patterns to arise in the narrative structure. Experiment 4 is evaluated as a whole and step by step in the design process (5 core steps). For experiments 1-3, we ran each for $500$ generations when using a pair of dimensions and for $250$ when using all dimensions. For experiment 4, each step in the design is done after $50$ generations; thus, we recorded data every $50$ generations.


\subsection{Metrics}

All our experiments are evaluated and analyzed following the same procedure and metrics, focusing on the novel generated individuals and their average across the $5$ runs. In particular, we focus on the \textit{average coverage}, \textit{average unique individuals}, \textit{average fitness}, and \textit{average interestingness}. \textit{Average coverage} is the cumulative coverage of the search space after a set of generations focused on the Step-Interestingness dimension pair. \textit{Average uniques} is a simple count of how many novel individuals were created throughout the experiments. \textit{Average fitness} calculates the average individual fitness in the search throughout all generations. Finally, \textit{average interestingness} calculates the average individual interestingness in the search throughout all generations.

\subsection{Narrative Graphs for Experiments}

Figure~\ref{fig:examples} and table~\ref{tab:level-constraints} show the target narrative graphs used in each experiment and their level design constraints, respectively. Experiments 1-3 use the proof-of-concept narrative graphs presented by Alvarez and Font~\cite{alvarez_tropetwist_2022} and experiment 4 uses a handmade narrative graph, exemplar of what a designer could create in Story Designer. 

Experiment 1 represents the overarching narrative structure of Super Mario Bros. (SMB)~\cite{nintendo_rd1_super_1985}. Mario (HERO) has as objective to rescue Princess Peach (HERO) from Bowser (BAD), who keeps Peach as a prisoner until Mario beats it, creating a derivative and conditional relation between Bowser and Peach. Before reaching Bowser, Mario must face "fake Bowsers" (DRA). Experiment 2 represents the structure from the eastern palace in \emph{Zelda: A Link to the Past} (Zelda:LttP)~\cite{nintendo_rd4_legend_1991}. Link (HERO) has as a goal the "Pendant of Courage" (MCG), but in order to collect it, Link must face ENEMY and BAD since there is a derivative pattern connecting them to (MCG). All palaces in \textit{A Link to the Past} follow a very similar structure and sequence. Experiment 3 represents a simplified overarching structure from \emph{Zelda: Ocarina of Time} (Zelda:OoT)~\cite{nintendo_rd4_legend_1998}. Young Link (HERO) has as a goal to collect/receive the Ocarina of Time (MCG), which enables the appearance of Adult Link (NEO). Achieving this goal creates conflicts between several heroes, Link and Zelda - Sheik (NEO, SH), and Gannondorf (BAD). 

Experiment 4 was designed with the capabilities of Story Designer in mind; step by step as a designer would create the narrative structure. First, it starts with the default structure; a HERO has a CONFLICT with an ENEMY. Subsequently, the structure is changed to fine-tune the ENEMY to BAD and create a goal (MCG) for the HERO. Another enemy (DRA) is added, creating a side conflict for the HERO (i.e., BAD is by definition the ``final boss"). Finally, the DRA is connected with the MCG with an entail connection, effectively making the DRA part of the game's main loop.

\section{Result and Analysis}

\begin{figure}[h]
    \centering
    \includegraphics[width=\linewidth]{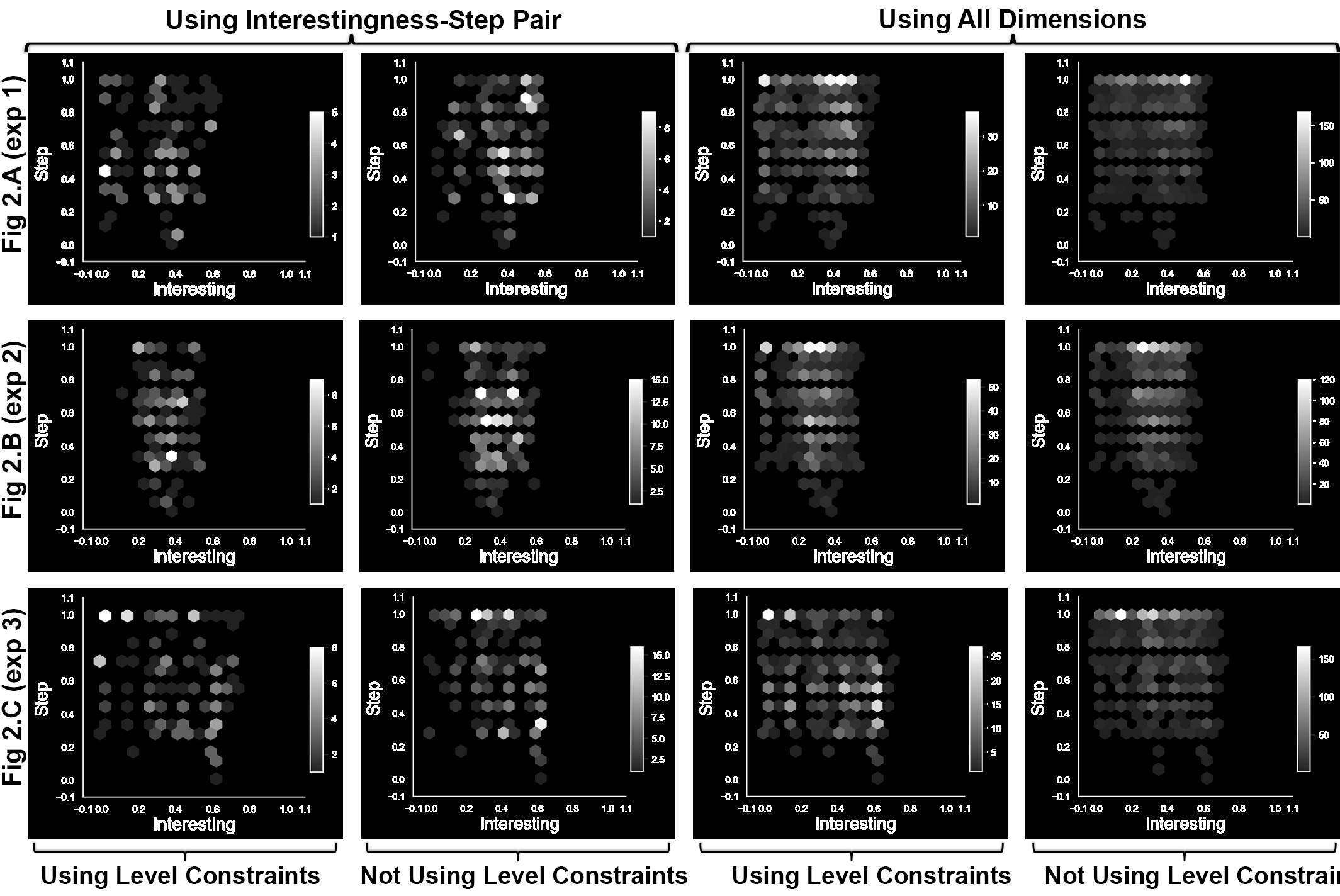}
    \caption{Rows show experiment 1-3, respectively. Each narrative graph can be seen in figure~\ref{fig:examples}.a-c, respectively. The first two columns are using interestingness-step as dimensions, and the other columns are using all dimensions in the search.}
    \label{fig:experiment123}
\end{figure}

\begin{figure}[t!]
    \centering
    \includegraphics[width=\linewidth]{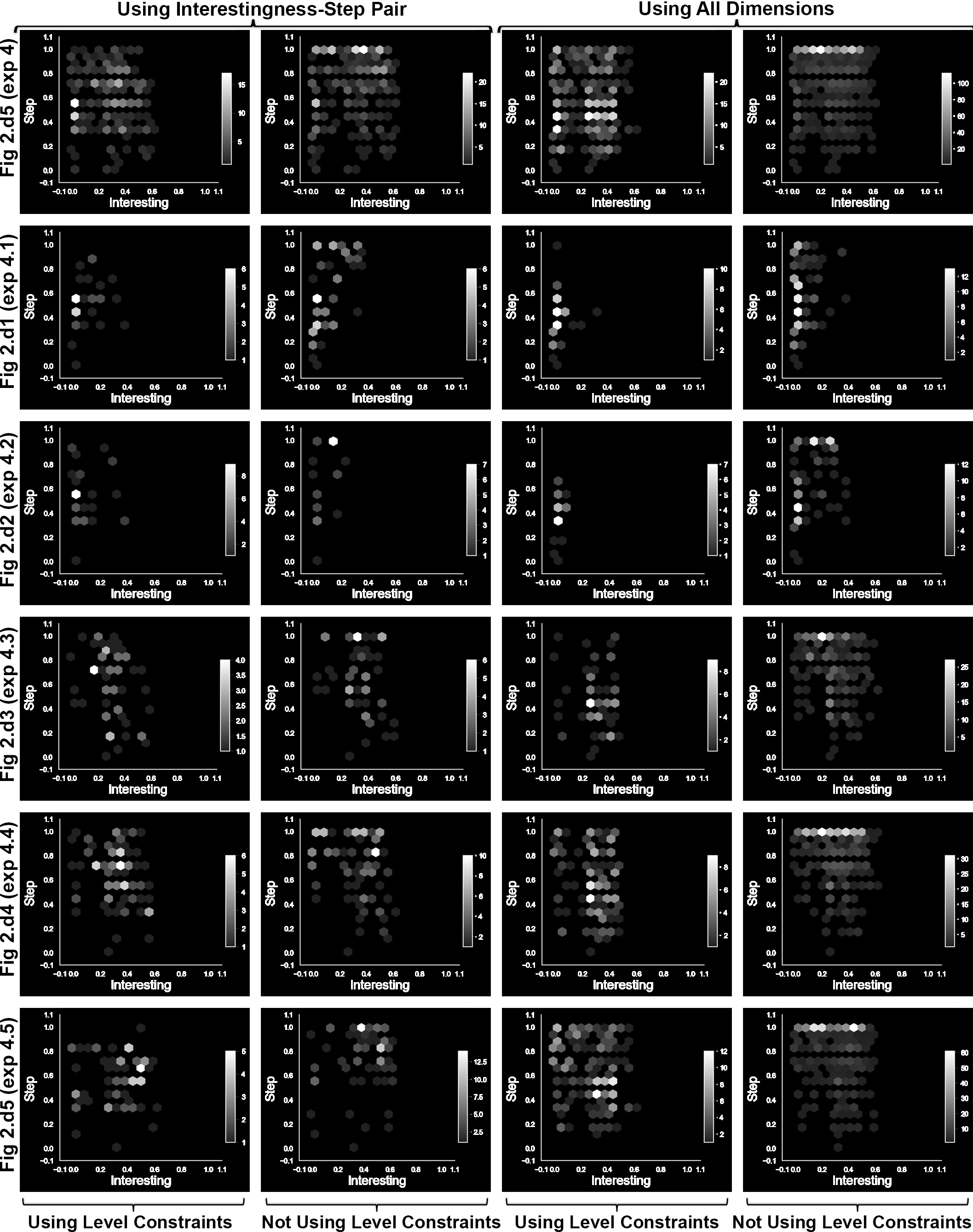}
    \caption{Expressive Range Analysis (ERA) and Temporal ERA (TERA) for Experiment 4. The first row is the general results for experiment 4 (ERA), while the subsequent ones are the individual results per step taken in the narrative structure design (figure~\ref{fig:examples}.d) (TERA). The first two columns are using interestingness-step as dimensions, and the other columns are using all dimensions in the search. }
    \label{fig:experiment-stepstep}
\end{figure}






In tables~\ref{tab:exp-int-step} and \ref{tab:exp-all-dims}, we present the results based on our metrics for the four experiments. Table~\ref{tab:exp-int-step} uses interestingness and step as dimension for MAP-Elites, while Table~\ref{tab:exp-all-dims} uses all dimension during search. To complement the analysis, figure~\ref{fig:experiment123} shows an exemplar expressive range analysis (ERA) for experiments 1-3 in the different configurations, and figure~\ref{fig:experiment-stepstep} shows an exemplar ERA for experiment 4 and an exemplar Temporal ERA (TERA) of the design steps. An ERA is an evaluation method to explore and visualize the expressiveness of an algorithm in content space~\cite{smith_analyzing_2010}. TERA is an extension of ERA that allows the inspection and analysis of changes in expressiveness over a defined period, which, when used in a non-aggregated fashion, as in experiment 4, shows the delta maps of the search~\cite{alvarez_assessing_2021}.

Analyzing and comparing the experiments show similar and consistent results across experiments regardless of using level design constraints or not, and using all dimensions or just a pair. Experiments 1-3 present consistent and stable results, similar among them in all metrics except coverage, which is more influenced by the specific graph and what type of information it provides, such as patterns, nodes, and connections. 

Experiment 4 shows MAP-Elites adaptability throughout the different design steps, especially visible in figure~\ref{fig:experiment-stepstep}. In the first two steps (4.1 and 4.2), MAP-Elites exploration is limited due to the narrative graph's simplicity. This is expected as the default narrative graph (HERO --> CONFLICT --> ENEMY) and the fine-tuned (i.e., ENEMY changed for BAD) has an interestingness score of 0 and, when used as a target, hinders the exploration with or without level constraints. However, as the design progresses, MAP-Elites adapt. Minimal input into the graph (experiment 4.3, onwards) improves the search and interestingness following the design's trend. IC MAP-Elites maintain properties such as adaptability and stability shown before for level design generation, making it adequate for the evolution of grammars and narrative structs as well. 


Experiment 4 also shows a concrete example of how the narrative graph would be used and designed by designers to change components in a game and enable different narrative structures. When put in context with the graphs for experiments 1-3, show relative diversity and expressiveness in the system. Experiment 4 and its steps show as well how the structure can relate to different "in-game" and level components, how, through the structure, designers can design main and side objectives, and how these could be approached. For instance, the DRA as a side conflict in the game and then incorporated as a main part of the game since to get the MCG, the HERO needs to face the DRA. That could then be used, in practice, to change, constrain, or adapt quests or part of the level design to be aligned with the structure.

Furthermore, both tables show similar patterns when using level design constraints or not. Fitness and interestingness vary slightly (avg. +0.009 and -0.004, respectively), whereas coverage and unique individuals are worst (avg. -3.1\%, -366.9, respectively).\footnote{These values are a combination of both table~\ref{tab:exp-int-step}, \ref{tab:exp-all-dims}} The lower unique individuals are expected since the search space is more constrained; thus, individuals that would otherwise be feasible (i.e., fully connected graph and without inadequate self-conflicts) would become unfeasible with the level constraints. A similar analysis could be expected from the slightly higher or comparable fitness since the lesser the individuals that are generated, the lesser the fitness variance. However, this also shows a practical and possible way to intertwine and enforce inter-facet constraints, since when adding level design constraints to the narrative generation, due to the more delimited space, the search can be more guided and focused and still generate quality-diverse content~\cite{gravina_constrained_2016,liapis_constrained_2015}.

The results point towards IC MAP-Elites, due to its constrained features and adaptability properties, being agnostic to these inter-facet constraints, which allows and ease the incorporation of these constraints in the system without having a major impact on the development. The tradeoff is then clearly that the possibility of the system to search for more or more complex narrative structures when using constraints is reduced since they would most probably violate constraints.

When comparing the use of a pair of dimensions (interestingness and step) and all dimensions in the search regardless of having or not constraints, the difference is expected regarding coverage (avg. 11.2\% more) and unique individuals (avg. 765.3 more) generation since MAP-Elites will be able to encounter and store elites in a bigger grid. However, the quality of the individuals is subpar in comparison with a pair of dimensions regarding fitness (avg. 0.08 more) and interestingness (avg. 0.04 more). This result is in line with~\cite{alvarez_interactive_2020}, where their results, applied to level design, showed more coverage and individuals generated when using all dimensions but focusing on suboptimal parts of the space. Figures~\ref{fig:experiment123} and~\ref{fig:experiment-stepstep} show that the experiments explore similar spaces, sparser when using a pair of dimensions and denser when using all dimensions. When observing the heatmap intensity, the search focus is distributed across the search space when using a pair of dimensions, while when using all dimensions, the search is focused on high step levels.

\section{Conclusions and Future Work}


In this paper, we have presented the first iteration of \emph{Story Designer} as a step towards a mixed-initiative co-creative system implementation of TropeTwist~\cite{alvarez_tropetwist_2022} to design narrative structures in EDD. The system allows the creation of narrative structures as narrative graphs that defines the overarching narrative, identifying characters, their roles and involvement, objectives, and core events. Story Designer also presents a step towards creating a holistic system, intertwining level design and narrative through simple level design constraints, effectively delimiting the search space of MAP-Elites with promising results. We analyzed and evaluated Story Designer and the impact of these level constraints through four simulated experiments; experiments 1-3 approach Story Designer in a more static scenario, while experiment 4 focuses on the step-by-step creation process.

Experiment 4 and, in general, the design process in Story Designer shows how the tropes, nodes, and connections, can be used to design a narrative structure step by step, changing the components of the narrative and how different elements in the game can be used and interpreted with simple changes. Defining conflicts among characters (thus, creating factions), primary and side objectives, as well as important elements in the narrative (e.g., plot devices), is a simple process. Changing these to adapt to the designer's goal is possible with minimal input. For instance, the change from experiment 4.4 to 4.5 (fig.~\ref{fig:examples}.d4,d5), where DRA passes from a side objective to a main part of the structure by creating an "entails" connection and forming a DerP meso-pattern (increasing the graph's interestingness score to 0.33). Equally important, the system preserves its properties and adapts to the created narrative graph, which could create a better experience for the designer. However, our evaluation was through simulated design sessions (especially, experiment 4) highlighting properties and tradeoffs of the system. We aim at evaluating Story Designer with a user study to assess its usability, the expressiveness designers have when creating structures, and the experience intertwining and creating level design constraints. 




Our next steps would be to continue the development of Story Designer to reincorporate the narrative structure into other facets and systems. Following a similar approach with constraints, narrative structures could constrain the search space for other facets, creating a feedback loop across facets and systems for a holistic approach. For instance, within EDD, narrative constraints could be reincorporated into both the level design facet~\cite{alvarez_empowering_2019} and the quest system to adapt main and side objectives~\cite{alvarez_questgram_2021}.

\bibliographystyle{ACM-Reference-Format}
\bibliography{references,auxiliary}

\end{document}